\documentclass{ws-p8-50x6-00}
\newcommand{\GeV}{\rm GeV}

\newcommand{\xg}{$x_\gamma$\,}
\newcommand{\Et}{$\overline{E}_t^2$\,}
\newcommand{\qq}{$Q^2$\,}
\newcommand{\EtEQ}{\overline{E}_t^2}
\newcommand{\gammaT}{$\gamma^*_T$\ }
\newcommand{\gammaL}{$\gamma^*_L$\ }

\begin{document}

\title{STRUCTURE OF VIRTUAL PHOTONS AT HERA~\footnote{T\lowercase{alk 
      given at} PHOTON 2001 C\lowercase{onference}, A\lowercase{scona},
    S\lowercase{witzerland}, S\lowercase{eptember} 2001}}

\author{K. SEDL\'AK~\footnote{S\lowercase{upported by} 
       GA AV\v{C}R \lowercase{grant} B1010005, 
       \lowercase{and projects} INGO-LA 116/2000,
       LN00A006}}

\address{Institute of Physics, AS CR, Na Slovance 2,
Praha 8, 182 21, Czech Republic\\E-mail: ksedlak@fzu.cz\\{\rm On 
behalf of the H1 and ZEUS Collaborations}}


\maketitle

\abstracts{
Triple differential dijet cross-sections in $e^\pm p$ interactions
measured with the H1 and ZEUS detectors at HERA are presented.
The data are compared to Monte Carlo
simulations which differ in their assumptions about photon structure
and parton evolution. Effects 
of the resolved
processes of longitudinally polarized virtual photons at HERA
are investigated for the first time.}

\section{Dijet Production at HERA}
\label{uvod}
The production of dijet events at HERA is dominated by processes,
in which a virtual photon, radiated from 
the electron, interacts with a parton in the proton.
In the region of photon virtuality $Q^2 \gg \Lambda_{\mathrm QCD}^2$, hard collisions
   of the photons do not necessitate the introduction of the concept
   of the resolved photon (as for the real photon) and the process can in
   principle be described by the direct photon contribution alone, in which 
   the photon interacts as a whole with partons from the proton.


The analyses presented here explore the region 
   $\Lambda_{\mathrm QCD}^2 \ll Q^2$ \raisebox{-0.8ex}{$\stackrel{\mathrm{<}}{\sim}$} 
$E_t^2$, where different theoretical approaches can
   be used to take into account higher order corrections.

\vspace{0.8em}
{\bf a) LO direct and resolved interactions} based on the DGLAP 
     evolution equations and parton showers allows
%
     the effects of transversally and
     also
     longitudinally polarized resolved photon 
     interactions to be studied~\cite{chyla,Chyla:2000hp}.
     Since the cross section for longitudinal photons vanishes 
     in the photoproduction
     regime due to gauge invariance, and 
     the concept of a resolved photon breaks down for $Q^2 > E_t^2$,
     the only evidence for this phenomena can be
     observed in the region $0 < Q^2 \ll E_t^2$.
     The main difference between the longitudinal ($\gamma^*_L$)
     and transverse ($\gamma^*_T$) virtual photon
     arises from the dependence of the respective fluxes on $y$. 
     While for $y\rightarrow 0$,
     both transverse and longitudinal fluxes are approximately same,
     the longitudinal flux vanishes for $y\rightarrow 1$.
     Also the dependence of the point-like (i.e. perturbatively calculable)
     part of the photon Parton Distribution Functions (PDF) 
     on \qq and $E_t^2$
     differs -- while the $\gamma^*_T$ PDF 
     is proportional to $\ln (E_t^2/Q^2)$, the
     $\gamma^*_L$ PDF depends on the scale in a typically hadron-like 
     manner.

\vspace{0.8em}
{\bf b) $\mathbf k_t$ unordered initial QCD cascades} accompanying the hard 
     process, employed for example in BFKL or CCFM evolution, 
     can lead to final 
     states in which the partons with the
     largest $k_t$ may come from the cascade, and not, as in DGLAP
     evolution, from the hard subprocess. Such 
     events may have a similar topology as is observed for the resolved 
     interactions introduced in the previous paragraph.
     This possibility is investigated using the CASCADE 
     generator~\cite{Cascade,Cascade2}
     based on the CCFM evolution equation.

\vspace{0.8em}
{\bf c) NLO calculations} with and without
     the concept
     of the resolved virtual photon offer another natural 
     possibility to be examined. Such comparisons are envisaged.

\vspace{0.8em}
All of these three approaches include higher order corrections
to the LO QCD matrix elements; however, each of the approaches treats them
differently.

\section{Measurement of Dijet Cross-Section}

The selection criteria defining the H1 and ZEUS dijet samples are
summarized in Table~\ref{tabulka}:
\begin{table}[ht]
\begin{center}
\begin{tabular}{|l|l|}
\hline 
{\bf H1} ~~ (16.3 pb$^{-1}$, 1999) & {\bf ZEUS} ~~ (38.2 pb$^{-1}$, 1996-97)\\ \hline 
$\sqrt{s}=318~\GeV$ & $\sqrt{s}=300~\GeV$ \\ \hline
$2~{\GeV}^2 < Q^2 < 80~{\GeV}^2$ & $0.1 < Q^2 < 10\,000~{\GeV}^2$ \\ \hline
$0.1 < y < 0.85$ & $0.2 < y < 0.55$ \\ \hline
$E_t^{jet\,1,2} > 5~\GeV$   &   $E_t^{jet\,1} > 7.5~\GeV$ \\
$\overline{E}_t > 6~\GeV$   &    $E_t^{jet\,2} > 6.5~\GeV$ \\ \hline
$-2.5 < \eta^{jet\,1,2} < 0$  &   $-3 < \eta^{jet\,1,2} < 0$ \\ \hline
\end{tabular}
%
%
%
\caption{Selection criteria of the dijet samples.}
\label{tabulka}
\end{center}
\end{table}
$Q^2$ denotes the photon virtuality, $y$ is the inelasticity,
$s$ is the total electron-proton centre-of-mass energy, 
$E_t^{\rm jet\,1,2}$ and
$\eta^{\rm jet\,1,2}$ are
the transverse energy and pseudorapidity of the jet with the highest 
or second highest
$E_t$, and $\overline{E_t}$ is defined as 
$(E_t^{\rm jet\,1}+E_t^{\rm jet\,2})/2$.

The measured data are corrected for detector effects using a bin-to-bin
method (ZEUS) or Bayesian unfolding (H1).  The largest source of
systematic
errors arises from the main calorimeter calibration uncertainty and, in
the case of H1, also from a model dependence of the detector correction.
The ZEUS measurement was presented in more detail at the
EPS 2001~\cite{Zeus}.

\begin{figure}[t]
\begin{center}
\epsfig{file=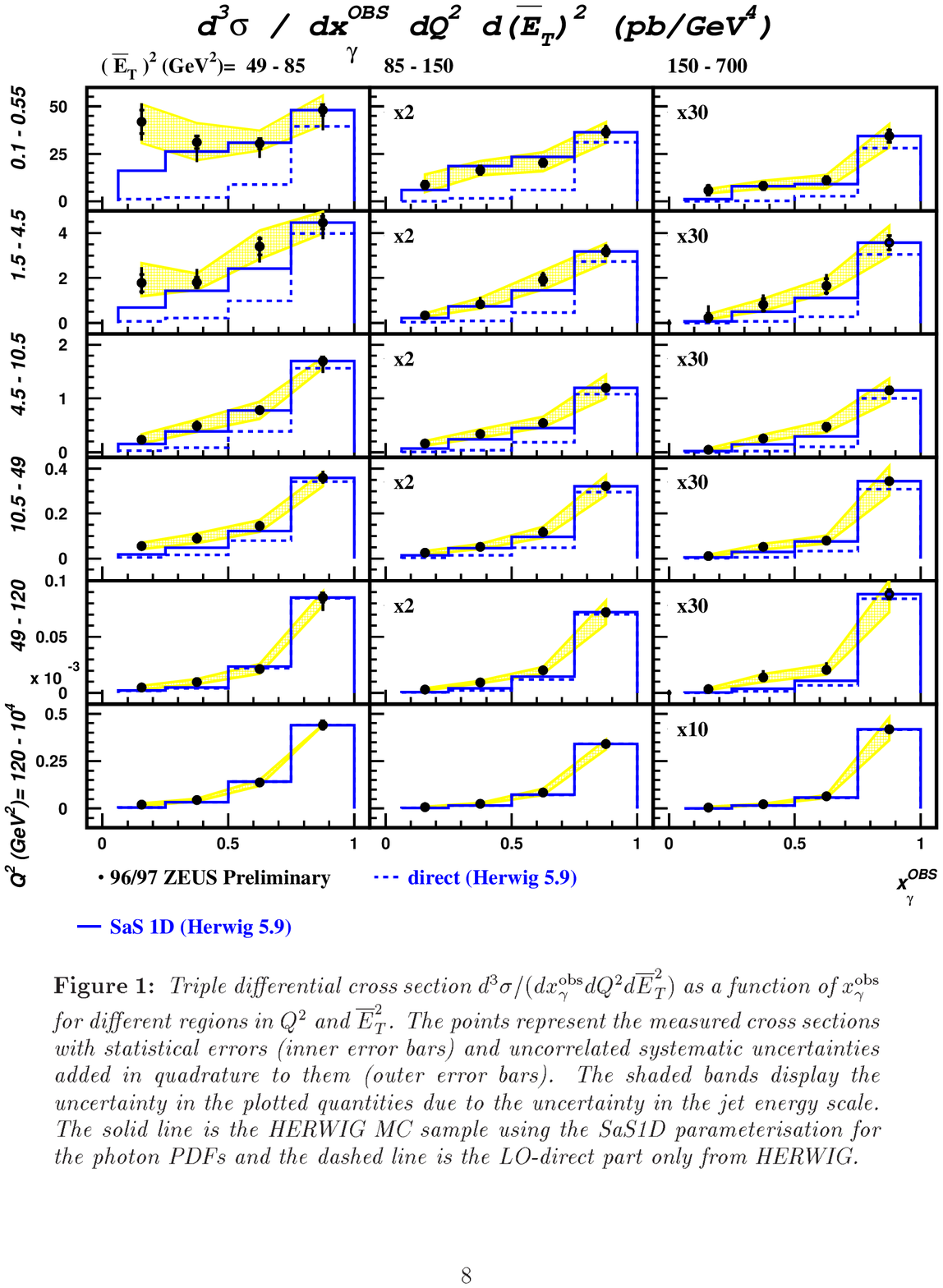,width=27pc,height=35pc,%
bbllx=45pt,bblly=229pt,bburx=526pt,bbury=710pt,clip=}
\caption{Triple differential dijet cross-section 
$d^3\sigma_{ep}/dQ^2 d\overline{E}_t^2 dx_\gamma$ for the ZEUS data
is depicted by points, the shaded bands display an uncertainty estimate
arising from the main calorimeter calibration.
The dashed histograms represent the direct part of HERWIG only, 
the full line stands for the sum of the direct and transverse resolved
contributions of HERWIG.}
\label{Zeus1}
\end{center}
\end{figure}
\begin{figure}[t]
\begin{center}
\epsfig{file=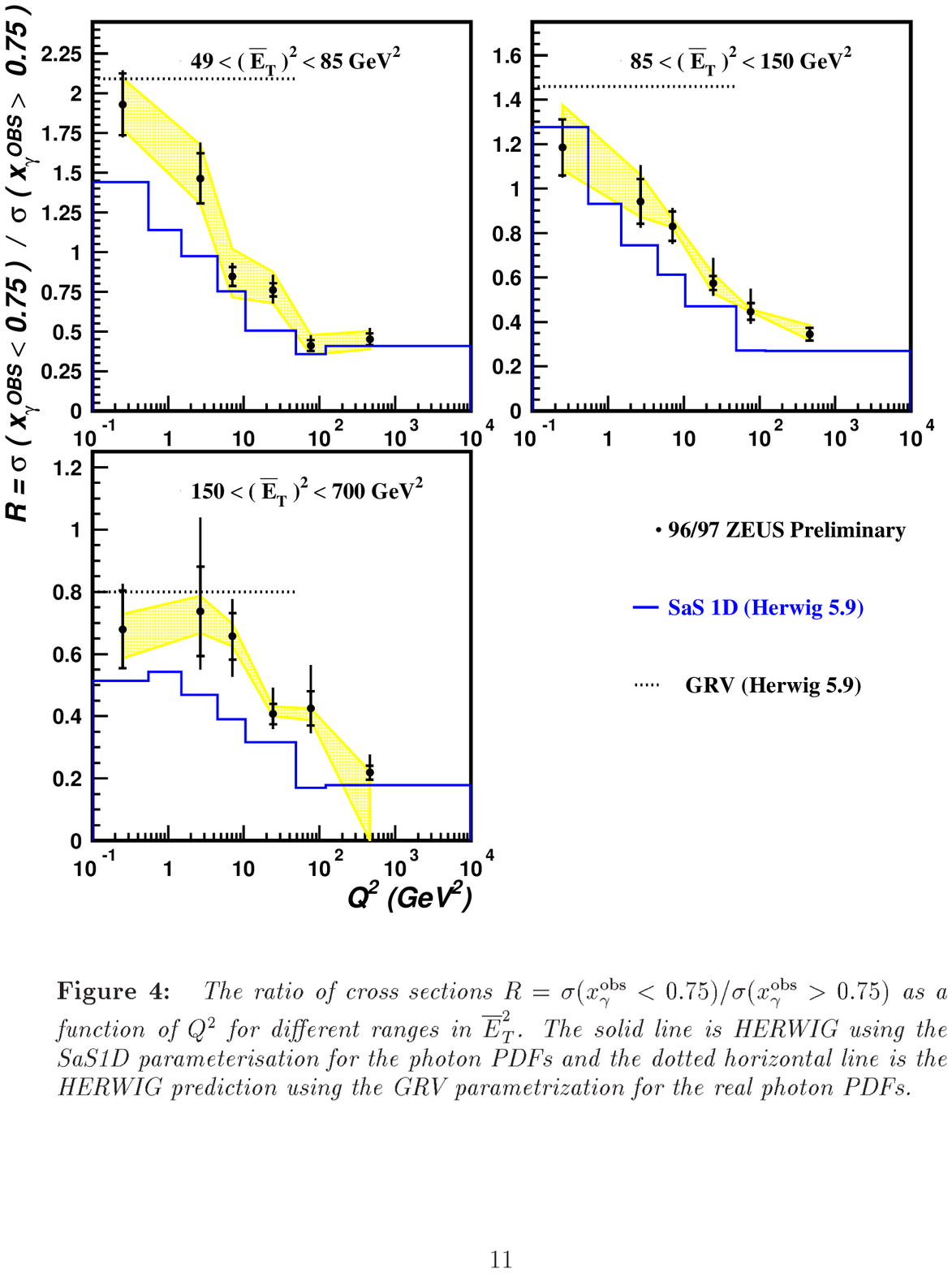,width=27pc,%
bbllx=75pt,bblly=218pt,bburx=520pt,bbury=655pt,clip=}
\caption{The ratio of cross-sections
$R=\sigma(x_\gamma < 0.75) / \sigma(x_\gamma > 0.75)$ from 
Fig.~\ref{Zeus1}.}
\label{Zeus2}
\end{center}
\end{figure}

\section{Results and Discussion}

The corrected triple-differential dijet cross-section measured
by ZEUS as a function of \qq, \Et and \xg is shown in
Fig.~\ref{Zeus1}.
A prediction of HERWIG with the SaS1D parameterization of 
the \gammaT
PDF, as well as the direct contribution of HERWIG
is compared to the data. Since the overall normalization of 
the LO Monte Carlo simulation is to some extent uncertain, the HERWIG
prediction has been normalized to the highest \xg bin $(x_\gamma > 0.75)$
in the data. The normalization is done separately for each 
$(Q^2,\EtEQ)$ bin.

In the region where $Q^2 > \EtEQ$, the data are well described by the
direct HERWIG component only.  
Resolved interactions are needed if $Q^2 < \EtEQ$.
However, even with the \gammaT resolved processes included,
HERWIG tends to underestimate 
the data in the lowest \qq region for $x_\gamma < 0.75$. 

This fact is also demonstrated in Fig.~\ref{Zeus2} by the ratio
of $\sigma(x_\gamma < 0.75) / \sigma(x_\gamma > 0.75)$. 
The slope of this distribution can be interpreted as 
a suppression of the virtual photon structure with increasing photon
virtuality.

\begin{center}
\begin{figure}[ht]
%
\epsfig{file=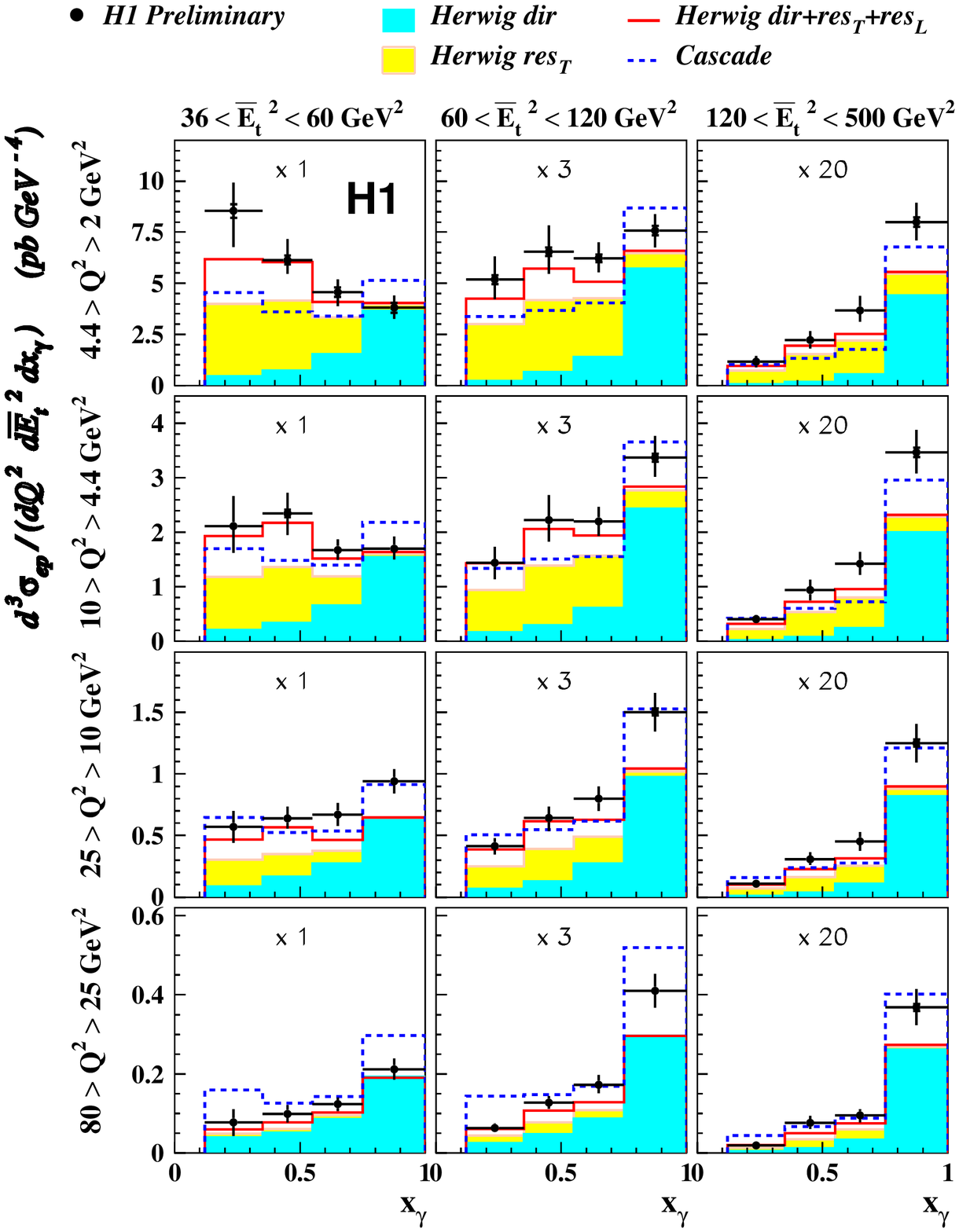,width=27.7pc,angle=0}

\caption{Triple differential dijet cross-section 
$d^3\sigma_{ep}/dQ^2 d\overline{E}_t^2 dx_\gamma$ for the H1 data
depicted by points is compared to predictions of the HERWIG and
CASCADE MC programs.\hspace{2cm}}
\label{H1}
\end{figure}
\end{center}

The corrected triple-differential dijet cross-section measured
at H1 as a function of \qq, \Et and \xg is shown in Fig.~\ref{H1}.
The H1 measurement is performed
in a different phase space (see Table~\ref{tabulka}) and Monte Carlo
predictions are not normalized to the data.

A comparison of the H1 measurement with HERWIG
leads to similar conclusions as 
drawn above
for ZEUS.  In addition, we can see that for the
highest \qq range ($25 < Q^2 < 80 ~\GeV^2$) and $x_\gamma<0.75$, 
the HERWIG direct contribution almost describes the data 
in the lowest \Et bin, but is too low in the highest \Et bin.  
This indicates an importance of the resolved processes
even at high \qq, once the hard scale, \Et, is high enough.

Standard HERWIG with direct and \gammaT resolved
contributions underestimates the data. The description 
is improved by adding \gammaL  resolved photon 
interactions,
which is done using a slightly
modified HERWIG with the correct longitudinal photon flux and 
a recent $\gamma^*_L$ PDF parameterization~\cite{Chyla:2000hp}.
As demonstrated in Fig.~\ref{H1}, the \gammaL resolved
contribution is significant, and brings  HERWIG
closer to the measurement.

\begin{center}
\begin{figure}
\epsfig{file=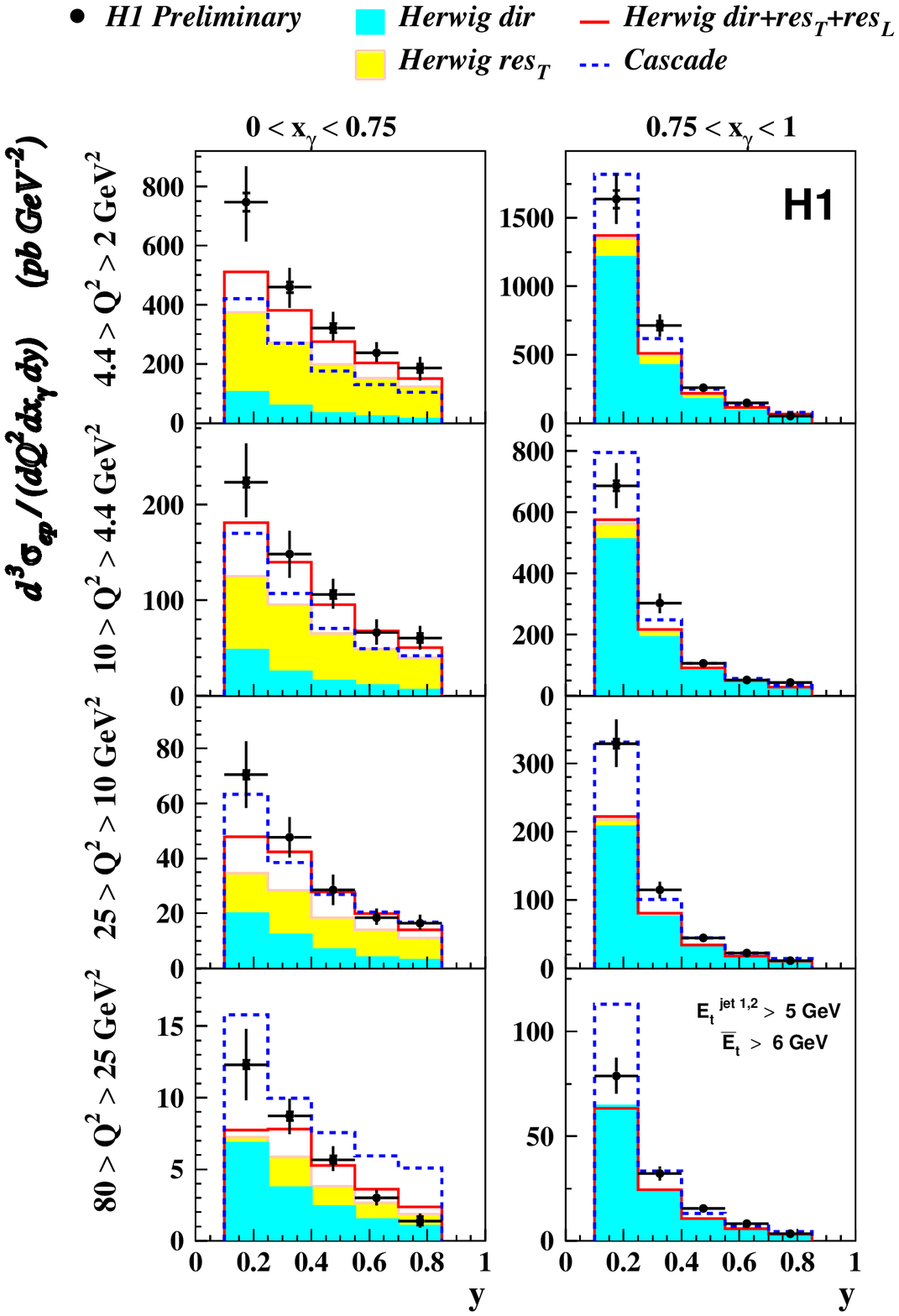,width=27.7pc,height=38pc,angle=0}
\caption{Triple differential dijet cross-section
$d^3\sigma_{ep}/dQ^2 dx_\gamma dy$.}
\label{H2}
\end{figure}
\end{center}

On the other hand, a simple enhancement of the PDF of the $\gamma^*_T$ 
in the 
resolved 
contribution could lead to a similar prediction
as the introduction of $\gamma^*_L$.
%
To eliminate this ambiguity,
the dijet cross-section has also been studied
as a function of \qq, \xg\, and $y$, which is shown in Fig.~\ref{H2}.
HERWIG is below the data, even if the resolved \gammaL
is added. 
This may be due to the uncertainty of the overall normalization of the
LO Monte Carlo prediction. 
In the region where $x_\gamma<0.75$, the slope of the HERWIG prediction 
depends significantly on whether $\gamma^*_L$ processes are included or
not. The $\gamma^*_L$ contributes significantly at low $y$,
while it becomes very small compared to $\gamma^*_T$ at high $y$. 
Unlike a pure enhancement
of $\gamma^*_T$ resolved processes by a constant factor,
the addition of $\gamma^*_L$ brings
the $y$ dependence of HERWIG much closer to the measurement.

As motivated in Section~\ref{uvod},
the measured cross-sections in Fig.~\ref{H1} are also compared 
to a prediction of the CASCADE MC program based on the CCFM evolution
approach. 
This theoretical concept does not involve 
any information about the virtual photon structure and involves
many fewer free parameters for tuning than the usual DGLAP-like MC
programs. Nevertheless, CASCADE describes the data well, 
except for the \qq dependence.
The \qq behavior, however, is related to the parameterization of
the unintegrated PDFs 
used in the program, which are not yet
constrained unambiguously.


\section{Conclusions}
The dijet cross-sections measured as a function of \qq, \Et, \xg\ and
\qq, \xg, $y$ at H1 and ZEUS have been presented.

  In the DGLAP evolution scheme, the importance of the \gammaT resolved
photon interactions is clearly demonstrated in the region where
$\EtEQ > Q^2$. Additional \gammaL resolved photon contributions
further improve the agreement of the HERWIG~5.9 prediction with the
measurement.

  Exploring the CCFM approach, the MC program CASCADE~1.0 gives
a qualitative description of the measured differential cross-sections;
however, the \qq dependence is not reproduced. 
On the other hand, 
the \xg dependence in CASCADE is comparable to the sum of the direct and
resolved contributions in DGLAP-like MC programs.


\end{document}